\documentclass[
    prl, twocolumn, superscriptaddress, nofootinbib, amsmath, amssymb,
    aps, floatfix
]{revtex4-2}

\usepackage{graphicx,subfigure}
\usepackage{svg}
\svgsetup{
    inkscapepath=i/svg-inkscape/
}
\svgpath{{svg/}}
\usepackage{xcolor}
\usepackage{setspace}
\usepackage{hyperref}

\usepackage[utf8]{inputenc}
\usepackage[T1]{fontenc}
\usepackage{lmodern}

\usepackage{booktabs}
\usepackage{color}
\usepackage{epsfig}
\usepackage{ifpdf}
\usepackage{amsmath}
\usepackage{bm}
\usepackage[english]{babel}
\usepackage{amsfonts}
\usepackage{amssymb}
\usepackage{braket}
\usepackage{enumerate}
\usepackage{cancel}
\usepackage{multirow}
\usepackage{cleveref}
\usepackage{xspace}
\usepackage{array}
\usepackage{fancyvrb}
\usepackage{fontawesome}
\usepackage{dcolumn}
\usepackage{slashed}
\usepackage[normalem]{ulem}
\usepackage{url}

\usepackage[absolute,overlay]{textpos} 
\def\XXint#1#2#3{{\setbox0=\hbox{$#1{#2#3}{\int}$}
     \vcenter{\hbox{$#2#3$}}\kern-.5\wd0}}

\makeatletter
\g@addto@macro\bfseries{\boldmath}
\makeatother

\definecolor{nicered}{rgb}{0.7,0.1,0.1}
\definecolor{nicegreen}{rgb}{0.1,0.5,0.1}
\hypersetup{colorlinks, urlcolor=blue, citecolor=nicegreen,linkcolor= nicered}

\begin{document}

\begin{textblock}{4}(0.2,0.1)   
  \includegraphics[width=2cm]{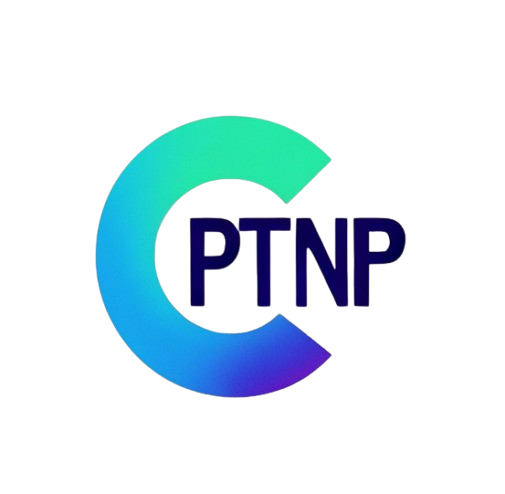}  
\end{textblock}

\begin{textblock}{6}(14,1)
  \raggedleft
  {\text{CPTNP-2025-002}}  
\end{textblock}


\title{Probing Millicharged Dark Matter with Magnetometer Coupled to Circuit}

\author{Yuanlin Gong}
\email{yuanlingong@nnu.edu.cn}
\affiliation{Department of Physics and Institute of Theoretical Physics, Nanjing Normal University, Nanjing, 210023, China}

\author{Hongliang Tian}
\email{hongliangtian@njnu.edu.cn}
\affiliation{Department of Physics and Institute of Theoretical Physics, Nanjing Normal University, Nanjing, 210023, China}

\author{Lei Wu}
\email{leiwu@njnu.edu.cn}
\affiliation{Department of Physics and Institute of Theoretical Physics, Nanjing Normal University, Nanjing, 210023, China}

\author{Bin Zhu}
\email{zhubin@mail.nankai.edu.cn}
\affiliation{School of Physics, Yantai University, Yantai 264005, China}

\begin{abstract}
We present a novel approach to detect millicharged dark matter (mDM) by using a high-sensitivity magnetometer coupled with the resonant and broadband readout circuits. In the external magnetic field, the interaction between mDM and the photon field introduces an effective current corresponding to the mDMs annihilation into photons that produces a faint oscillating magnetic field signal, with a frequency uniquely determined by twice the mDM mass. By calculating the expected signal for two experimental configurations -- toroidal and solenoidal magnetic fields -- we show the potential to explore the uncharted regions of mDM parameter space. Our analysis establishes unprecedented constraints on the mDM coupling constant across the mass range $1\times 10^{-12}~\mathrm{eV}$ to $6 \times 10^{-8}~\mathrm{eV}$, surpassing existing experimental limits by more than ten orders of magnitude.

\end{abstract}

\maketitle
\newpage

{\it Introduction} --- The secret of the dark matter continually provokes the pursuit in particle physics, astrophysics, and cosmology \cite{Bertone:2004pz,Cirelli:2024ssz}. The luminance of dark matter are proved to be faint by plenty of the observations. But particles with tiny millicharge under electromagnetism are still feasible dark matter candidates \cite{Goldberg:1986nk,Brahm:1989jh,Dimopoulos:1989hk,DeRujula:1989fe,Feldman:2007wj,McDermott:2010pa,Chu:2011be,Cline:2012is}. The conspicuous anomaly of the 21cm signal reported by the EDGES Collaboration can be well explained in the presence of the millicharged dark matter (mDM) \cite{Munoz:2018pzp,Berlin:2018sjs,Slatyer:2018aqg,Kovetz:2018zan,Liu:2019knx}. More recently, the constraint on the charge radius of mDM is derived by the PandaX-4T experiment \cite{PandaX:2023toi} with its nuclei recoil detector. There are also many minimal dark sector extensions \cite{Holdom:1985ag, Goldberg:1986nk,Izaguirre:2015eya,Feldman:2007wj,Cheung:2007ut,Feng:2023ubl}, string theory compactifications \cite{Wen:1985qj,Burgess:2008ri, Goodsell:2009xc, Cicoli:2011yh, Shiu:2013wxa, Feng:2014eja}, and grand unification theories \cite{Pati:1973uk, Georgi:1974my, Preskill:1984gd} that mDM can naturally arise. The freeze-in \cite{Hall:2009bx,Dvorkin:2019zdi,Bhattiprolu:2023akk,Bhattiprolu:2024dmh} or misalignment mechanism \cite{Preskill:1982cy,Abbott:1982af,Dine:1982ah,Nelson:2011sf,Arias:2012az} can support the production of the mDM in the early Universe and generate proper relic density today.

Scalar mDM can be ultra-light and act as an alliance member of the Ultra-Light Dark Matter (ULDM), a mass features macroscopic quantum fluctuations. It provides a natural solution to the small-scale structure problems~\cite{Hu:2000ke, Hui:2016ltb, Ferreira:2020fam} and receives growing interest. The early Universe behaviors of millicharged ULDM was studied with particular focuses on misalignment production \cite{Bogorad:2021uew} and cosmological stability \cite{Jaeckel:2021xyo}. Astrophysical and cosmological observations, including stellar evolution \cite{Bernstein:1963qh,Dobroliubov:1989mr,Davidson:1991si,Davidson:2000hf,Vogel:2013raa,Fung:2023euv}, SN1987A \cite{Mohapatra:1990vq,Fiorillo:2024upk}, timing of radio waves \cite{Caputo:2019tms}, and  so on \cite{Davidson:1993sj,Melchiorri:2007sq,Burrage:2009yz,Ahlers:2009kh,Jaeckel:2010ni,Vinyoles:2015khy} can impose strong limits on the coupling $e_m$ between the mDMs and the photons in the sub-eV mass range. For the terrestrial searches, plenty of proposals and experiments, cavities filled with strong electric field to probe the electric current produced from the Schwinger pair production \cite{Gies:2006hv, Berlin:2020pey,Romanenko:2023irv}, a magnetic field signal in the geomagnetic filed environment~\cite{Arza:2025cou}, observation of the invisible decays of positronium \cite{Badertscher:2006fm}, lamb shift of the hydrogen atom \cite{Gluck:2007ia}, laser polarization experiments \cite{Gies:2006ca,Ahlers:2007qf,DellaValle:2014xoa, DellaValle:2015xxa}, and Cavendish experiment for testing Coulomb's Law \cite{Jaeckel:2009dh} have imposed robust constraints. 

We note that the interaction between mDM and the photon field in an external electromagnetic field introduces an effective current, which originates from the annihilation of mDM into photons. For the Compton wavelength of a mDM exceeding the size of experimental apparatus, the generated signal appears as a quasi-static magnetic field signal. Unlike the axion and dark photon cases \cite{Sikivie:2013laa,Chaudhuri:2014dla,Arias:2014ela,Kahn:2016aff,Chaudhuri:2014dla,Fedderke:2021aqo,Arza:2021ekq,Sulai:2023zqw}, the signal of mDM is inversely proportional to the square of the mDM mass $m_{\phi}$, which makes our detection method highly sensitive to ultra-light mDM.

Inspired by this feature, we propose a detection method for mDM by using a high-sensitivity magnetometer coupled with the resonant and broadband readout circuits. We also find that the direction of the effective current in mDM is determined by the vector potential of the magnet, rather than the magnetic field itself. This leads to a differently oriented signal, necessitating distinct pickup loop geometries and relative positions for the same magnet configuration compared to those in axion and dark photon scenarios \cite{Sikivie:2013laa,Chaudhuri:2014dla,Arias:2014ela,Kahn:2016aff}. We consider the toroidal and solenoidal magnetic fields, which are relatively regular, facilitating precise theoretical calculations and experimental operations. Leveraging the above-mentioned advantages, we calculate the expected signals and obtain the unprecedented constraints on the mDM coupling in the mass range from $1\times10^{-12}$ eV to $6\times10^{-8}$ eV, exceeding existing experimental limits by more than ten orders of magnitude. This demonstrates great potential for exploring uncharted regions of the mDM parameter space and will open up a new direction for mDM.

\begin{figure*}[ht]
\centering
\subfigure{
\includegraphics[width=0.5\linewidth]{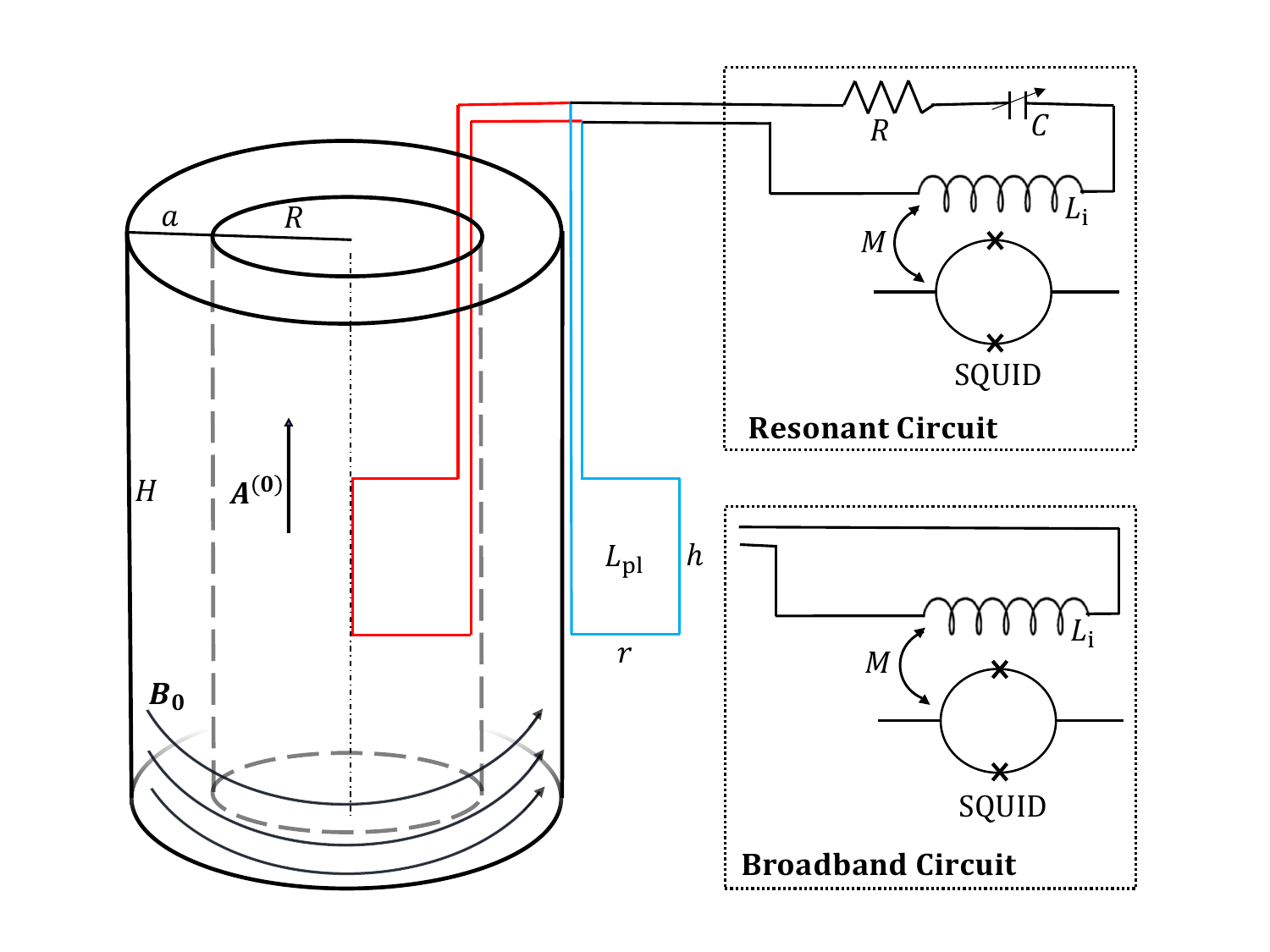}
\includegraphics[width=0.5\linewidth]{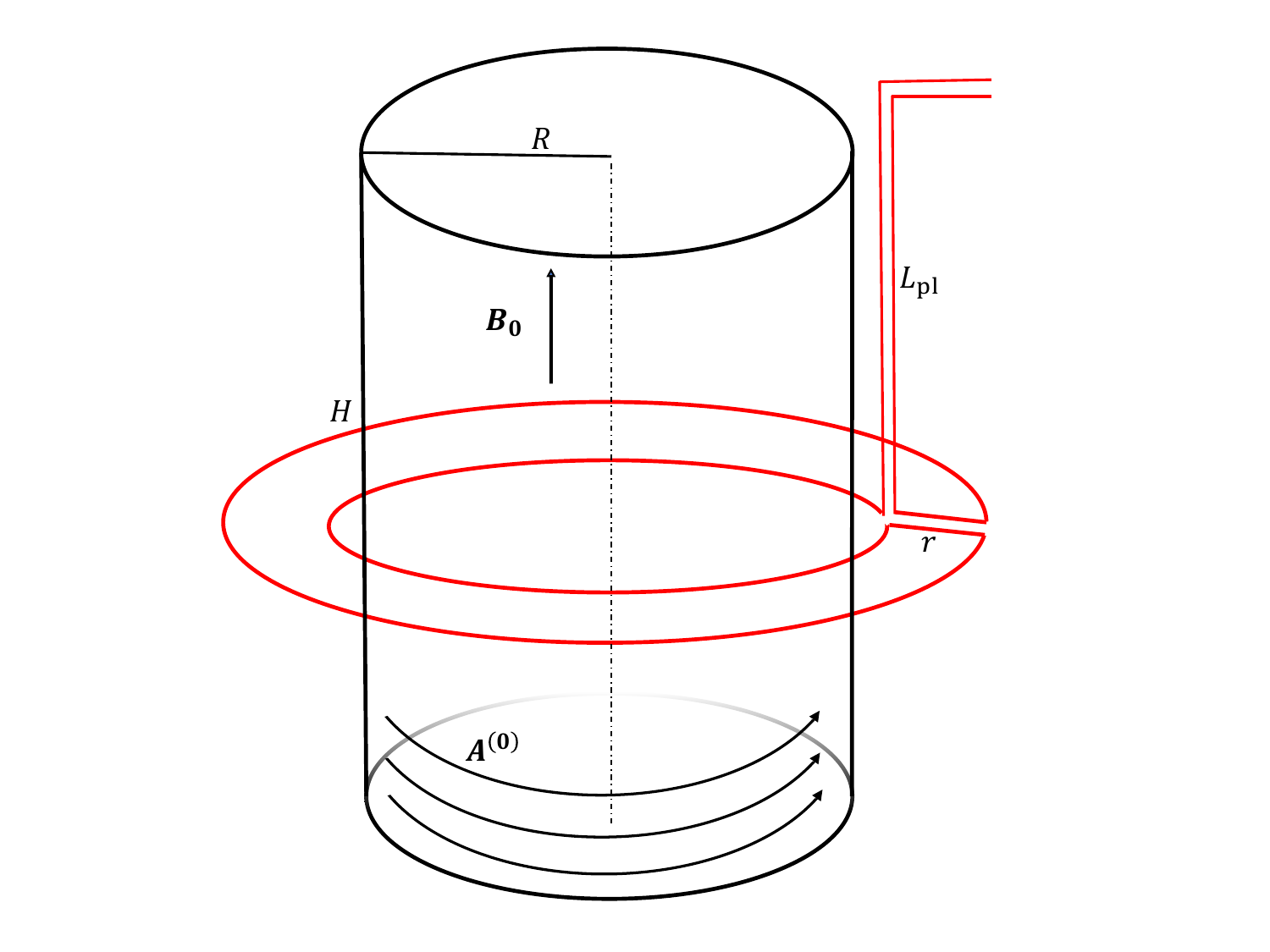}
}
\caption{Schematic diagram of the detector proposed to detect the millicharged Dark Matter (mDM). \textbf{Left:} A cylindrical toroid magnet (with height $H$, inner radius $R$, and width $a$) produces a confined magnetic filed $\mathbf{B}_0$ and unconfined vector potential $\mathbf{A}^{(0)}$. In the background of the $\mathbf{A}^{(0)}$, mDM sources a effective current in the same direction and therefore an axial magnetic field, which is measured by the superconducting pickup loop (with width $r$ and height $h$) linked to readout circuit (See text for details). \textbf{Right:} Different from the former, here the magnet is a solenoid (with radius $R$ and height $H$) and the mDM sourced magnetic filed is along the $Z$-axis. A ring pickup loop (with inner radius $R$ and outer radius $R+r$) centered in the some direction is used to detect the magnetic filed. }
\label{setup}
\end{figure*}

{\it Electrodynamics modified by Millicharged Dark Matter} --- This work focuses on ultralight scalar dark matter $\phi$ effectively charged under the $U(1)_{\rm{em}}$ with a millicharge $e_m$, remaining agnostic about the origin of the interaction. The coupling between $\phi$ and the photon field $A_\mu$ naturally emerges from gauge invariance rather than introduced through an ad-hoc interaction, 
\begin{align}
{\cal L}=D_\mu\phi(D^\mu\phi)^*, \label{eq:lag1}
\end{align}
with $D_\mu=\partial_\mu+ie_mA_\mu$ the covariant derivative. Under the Coulomb gauge $\vec\nabla\cdot\vec A=0$ and $A_0=0$, in an external electromagnetic field $A_\mu^{(0)}=(A^{(0)}_0,\vec A^{(0)})$, the mDM modified Maxwell Equations at leading order in $e_m$ are shown as \cite{Arza:2025cou},
\begin{eqnarray}
\nabla\cdot \vec E &=&2e_m\phi^{(0)}\text{Im}(\partial_t \phi^{(1)}), \label{eq:max1}\\
\nabla\times\vec E+\partial_t{\vec B}&=&0, \label{eq:max2} \\
\nabla\cdot \vec B&=&0 , \label{eq:max3}\\
\nabla\times \vec B-\partial_t{\vec E}&=&-2e_m^2A^{(0)\nu}|\phi^{(0)}|^2. \label{eq:max4}
\end{eqnarray}
Here we have perturbatively expand $\phi$ as $\phi=\phi^{(0)}+\phi^{(1)}$. The $\phi^{(0)}=\phi_0\cos(\vec k_\phi\cdot\vec x-m_\phi t)$ is the scalar field free of any interaction to ensure charge symmetry. The $k_\phi$ is the momentum of the mDM and $\phi_0=\sqrt{2\rho}/m_\phi$, with $m_\phi$ the mass of the mDM, and $\rho=0.3\,\text{GeV}/\text{cm}^3$ being the local mDM energy density. The field $\phi^{(1)}$ captures the first order perturbation (in the very small coupling $e_m$) to the $\phi^{(0)}$ in the presence of external field $A^{(0)}_\mu$ and is necessarily introduced in to preserve the gauge invariance \cite{Arza:2025cou}. It is determined by ${(\Box+m_\phi^2)} \phi^{(1)}=-2ie_m\vec\nabla\phi^{(0)}\cdot\vec A^{(0)},$
with the boundary condition $\phi^{(1)}=0$ for $|\vec x|\rightarrow\infty$.

The $\vec J_\text{eff}=-2e_m^2\vec A^{(0)}|\phi^{(0)}|^2$ in Eq. (\ref{eq:max4}) is the effective current that explicitly determined by the external vector potential. We can derive the vector potential $\vec A^{(0)}$ by resorting to $\vec B_0=\nabla\times\vec A^{(0)}$ in a given external magnetic filed $\vec B_0$. In the magneto quasi static limit, i.e., for mDM with Compton wave
length bigger than experiment length scale, the effective current can drive a magnetic field signal $\vec B$, oscillating at frequency $2 m_\phi$,  being perpendicular to $\vec A^{(0)}$ and parallel to the $\vec B_0$.

We propose to search the magnetic field signal in the magnetic field background of the toroid and solenoid with resonant and broadband readout circuits as shown in Fig. \ref{setup}. In the presence of the mDM, the magnets will act as the effective currents by virtue of their vector potential fields and generate the magnetic fields signal $\vec B$ passing through the pickup loops placed in the zero magnetic filed region. We note that, there is a gap for the toroid magnet not shown. It provides the complete circuit that Meissner current can return along the outside surface of the toroid \cite{Kahn:2016aff}. Also, we use two rectangular pickup loops, one hanging in the middle of the toroid hole, the other nearly outside the toroid skeleton working together to detect the signal. While for the solenoid, the ring pickup loop is hanging nearly outside the magnet skeleton. These two common magnets are previously utilized to search for the axion, dark photon, and ultrlight scalar dark matter \cite{Sikivie:2013laa,Chaudhuri:2014dla,Kahn:2016aff,Crisosto:2019fcj,Donohue:2021jbv,Zhang:2021bpa,Salemi:2021gck}.

In particularly, we note that there are two differences with respect to the axion case. Own to the direction dependence on the vector potential but not on the magnetic field of the effective current, the magnetic field signal pours out in a different direction. It necessitates the different pickup loop geometry and relative position in the same magnet configuration. Secondly, attribute to the two body annihilation of the mDM, we find the effective current $\vec J_\text{eff}\propto |\phi^{(0)}|^2$. While the effective current induced by axion decay is $\vec J_\text{a} = g_{a\gamma}\vec{B}_0\partial_ta$, which is proportional to the axion amplitude $a_0=\sqrt{2 \rho}/m_a$. Here, $g_{a\gamma}$ is the coupling between axion and two photons.
Given $\phi_0=\sqrt{2 \rho}/m_\phi$, in the magneto quasi static limit, the induced magnetic field signal scales as $m_{\phi}^{-2}$ distinct from the axion case, where the signal is mass-independent. It is this fact that makes the experiment proposed be very sensitive to the lower mass range and have different scaling on the energy density $\rho$ compared to the axion case.

{\it Magnetic field flux of signal} --- We consider the signal from the toroid of rectangular cross section (height $H$, width $a$) and inner radius $R$, as illustrated in Fig. \ref{setup}. For the magnetic field 
\begin{equation}
    \vec B_0(\rho)= \frac{B_0 R}{\rho}\hat e_\phi, \quad R<\rho<R+a,
\end{equation}
confined inside the toroid volume, in the Coulomb gauge, we can find the corresponding vector potential as \cite{Carron1995}
\begin{align}
    \vec A^{(0)}(\rho)=B_0 R \hat e_z \begin{cases}
        \ln{(\frac{a+R}{R})},& 0<\rho<R\\
        \ln{(\frac{a+R}{\rho})},& R<\rho<R+a\\
        0, & R+a<\rho
    \end{cases},
\end{align}
where the $B_0$ is the magnitude of $\vec B_0$ at the inner radius. Here, for simplicity, we worked in the high toroid limit $H\gg a,R,r,h$, with pickup loop width $r$ and height $h$. For a general toroid, there is a component for $\Vec{A}_0$ lies in $\hat \rho$ direction that will source a magnetic field anti-parallel to that from the $\hat z$ direction and thus reduces the signal. In the magneto quasi static limit, $m_\phi H\ll 1$, the displacement current in Eq. (\ref{eq:max4}) do not work. Therefore, by virtue of the Biot-Savart Law, the flux of $\vec B$ through the pickup loop in the toroid hole can be integrated out as
\begin{align}
    \Phi_{\text{pl,1}}=\frac{1}{2}e_m^2\phi_0^2 B_0 R h r^2 \ln(\frac{a+R}{R})\cos(2m_\phi t),
\end{align}
along with the angular frequency given
by twice the mDM mass and the bandwidth $\Delta \omega\approx 10^{-6} m_\phi$. For the second pickup loop closely hanging outside the toroid, the flux is given by
\begin{align}
    \Phi_{\text{pl,2}}=\frac{1}{2}e_m^2\phi_0^2 B_0 R h a(a+2R)\ln(\frac{a+R+r}{a+R})\cos(2m_\phi t).
\end{align}
The total magnetic flux in the pickup loops is $\Phi_{\text{pl}}=\Phi_{\text{pl,1}}+\Phi_{\text{pl,2}}$. The receiving rate of the pickup loop magnetic energy of self-induction gives the signal power of the pickup loop as $P=m_\phi|\Phi_\text{pl}|^2/L_\text{pl}$, where $L_\text{pl}$ is the self-inductance of the pickup loop. For a rectangular pickup loop with length $h$, width $r$ and loop radius $\kappa$, the self-inductance is estimated to be $L_{\rm{pl}}\approx \frac{h}{\pi}\ln(\frac{r}{\kappa})$.

\begin{figure*}
    \centering
    \includegraphics[width=0.8\linewidth]{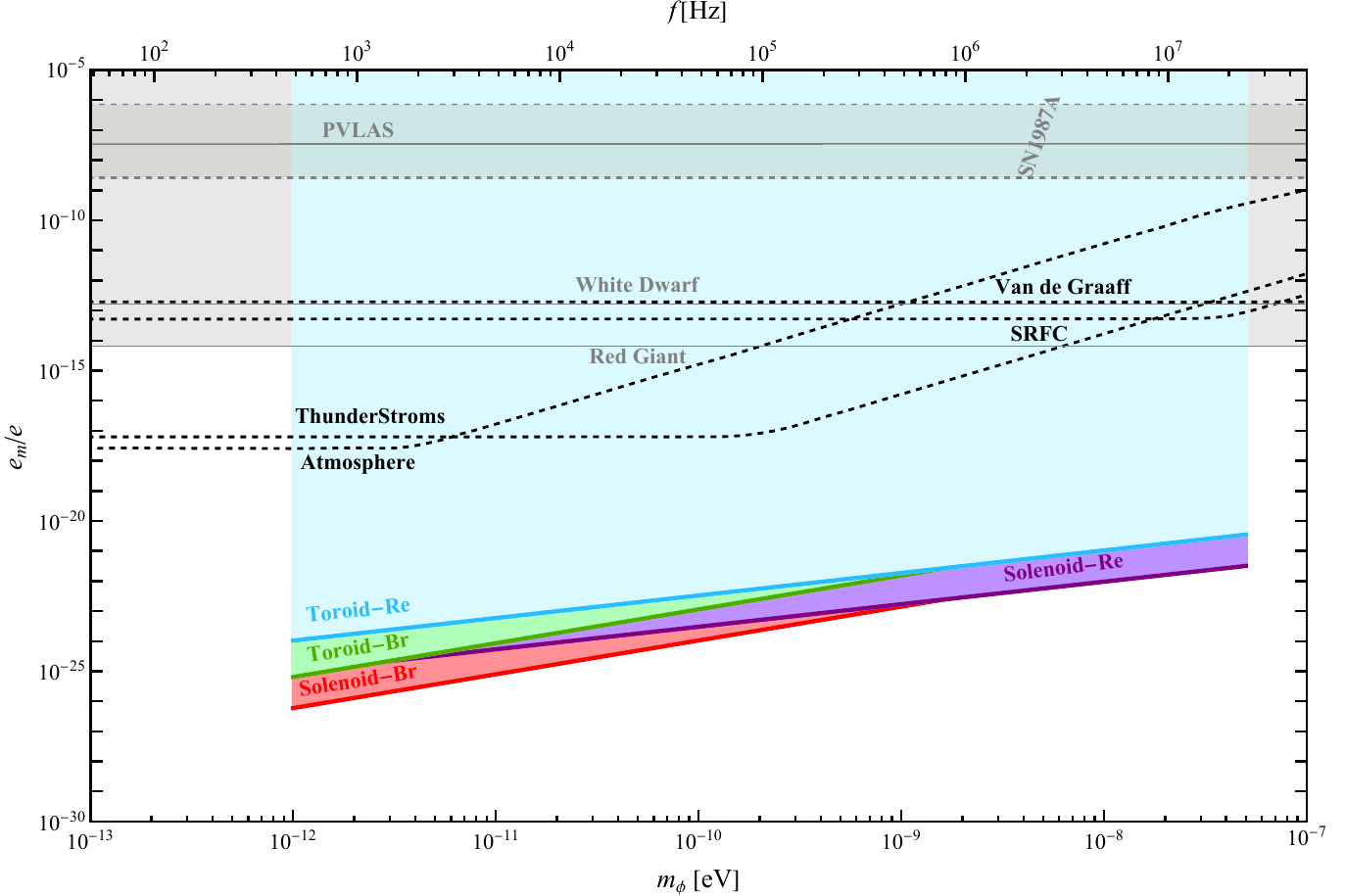}
    \caption{Projected limits on the $e_m/e$ in the plane $e_m/e$ verse $m_\phi$ ($e$ is the elementary electric charge). The blue, green, purple and red shaded regions are excluded by the resonant and broadband detection with toroid and solenoid magnets. We plot the best existing experiment constraints from the PVLAS experiment \cite{DellaValle:2015xxa}. Previous astrophysical and cosmological observation bounds (gray lines) including SN1987A \cite{Fiorillo:2024upk},  Red Giant \cite{Fung:2023euv}, and White Dwarf Cooling \cite{Davidson:2000hf} are also shown. The dashed black lines are projected limits from Superconducting Radio-Frequency Cavities \cite{Berlin:2020pey} and millicharged Condensates (including ThunderStorm, Atmosphere, and Van de Graaff cases) \cite{Berlin:2024dwg} in the recently studies.}
    \label{fig:limit}
\end{figure*}

On the other hand, the uniform magnetic field $\vec B_0=B_0\hat e_z$ in the solenoid magnet give the following vector potential
\begin{align}
    \vec A^{(0)}(\rho)=\frac{1}{2}B_0  \hat e_\phi \begin{cases}
        \rho,& 0<\rho<R\\
        \frac{R^2}{\rho},& R<\rho<\infty
    \end{cases}.
\end{align}
There are also ``fringing fields'' outside the solenoid bore that are neglected in our preliminary analyse. Maybe we can shield these field with superconducting material. The flux received by the ring pickup loop reads
\begin{align}
   {\varPhi}_{\text{pl}}=\frac{1}{2}e_m^2\phi_0^2 B_0 R V_A\cos(2m_\phi t), 
\end{align}
with the geometry factor $V_A$ is explicitly given by
\begin{align} V_A=&\int_{R}^{R+r}\mathrm{d}\rho^{\prime}\int_0^{R}\mathrm{d}\rho\int_0^{2\pi}\mathrm{d}\theta'\:\frac{H\rho^2\rho^{\prime}(\rho-\rho^{\prime}\cos\theta')}{R\tilde{\rho}^2\sqrt{H^2+4\tilde{\rho}^2}}\nonumber\\
+&\int_{R}^{R+r}\mathrm{d}\rho^{\prime}\int_{R}^{\infty}\mathrm{d}\rho\int_0^{2\pi}\mathrm{d}\theta'\:\frac{HR\rho^{\prime}(\rho-\rho^{\prime}\cos\theta')}{\tilde{\rho}^2\sqrt{H^2+4\tilde{\rho}^2}},
\end{align}
where $\tilde{\rho}^2\equiv \rho^2+\rho^{\prime2}-2\rho\rho^{\prime}\cos\theta'$. Correspondingly, the self-inductance of the ring pickup loop is taken as $L_{\rm pl} = \frac{(R+r)}{2}(\ln\left(\frac{\xi \pi(R^2+2Rr)}{\kappa^2}\right)+\frac{1}{2})$, with $\xi$ a factor of unity. Given $R,\,r,\,H$, we can evaluate $V_A$ numerically. To make a sensitivity estimation, we set $R=a=r=h=H/10=0.1\, \text{meter}, \,1\, \text{meter}$ as two stages of the experiment. These give the $V_A= 2.9 \,\mathrm{m^3}$ and $152.7 \,\mathrm{m^3}$. Also, the magnetoquasistatic limit restricts the upper limit on the mass range of interest as $\sim 6\times 10^{-8}\,\mathrm{eV}$ and $\sim 6\times 10^{-9}\,\mathrm{eV}$ in the two stages correspondingly.

{\it Detection Methods} --- We firstly consider an LC circuit to amplify the $\Phi_{\text{pl}}$ by the quality factor $Q$ at the resonant frequency and subsequently detect the amplified flux with a Superconducting Quantum Interference Device (SQUID) magnetometer. As shown in left plane of the Fig. \ref{setup}, the LC circuit  bridges the pickup loop $L_\text{pl}$ with the SQUID magnetometer $L_\text{m}$. And its probe coil $L_\text{i}$ inductively coupled to the SQUID magnetometer with mutual inductance $M$ to services as a tuned magnetometer through a tunable capacitor $C$. The total inductance $L=L_\text{pl}+L_\text{m}+L_\text{i}$ is dominated by the pickup loop due to its much larger geometry area. For a finite capacitance, the LC circuit resonates at frequency $\omega = 1/\sqrt{LC}$ with quality factor $Q=\omega/\Delta\omega_\text{LC}=(\omega C R)^{-1}$, where $\Delta\omega_\text{LC}$ is the bandwidth of the LC circuit and $R$ is the intrinsic resistance of the capacitor. In each tune of the LC circuit, the signal oscillating at $\omega$ with bandwidth $\Delta\omega$ is searched. High quality factor can be achieved by advanced resonant circuit \cite{403276,Myers2007,Kahn:2016aff} or superconducting wire for the part of the LC circuit \cite{Devlin:2021fpq}. Be conservatively, We take $Q=10^4$ as a benchmark. The enhanced signal power received by the SQUID magnetometer writes as $P_\text{Re}=m_\phi|\Phi_\text{Re}|^2/L$ with the resonant amplified magnetic flux being $\Phi_\text{Re}=QML^{-1}\Phi_\text{pl}$ \cite{Sikivie:2013laa,Crisosto:2019fcj}. As a result, the signal to noise ratio at resonance is calculated from the Dicke radiometer equation \cite{1946RScI...17..268D}
\begin{equation}
    \frac{S}{N}=\frac{P_\text{Re}}{kT} \sqrt{\frac{\Delta t}{\Delta \omega_\text{LC}}},
\end{equation}
with where $k$ is Boltzmann's constant, $T$ is the environment temperature of the readout system, and $\Delta t$ is integration time of each bandwidth search. Precisely, the noise comes from the SQUID magnetometer with contributions from thermal fluctuations of both voltage and current and the LC circuit. The thermal noise of the circuit resistance dominant the total noise for conventional LC circuit, while that of a novel LC circuit design receives main contribution from the magnetometer thermal noise \cite{Devlin:2021fpq,Zhang:2021bpa}. Following them, we take $T= 4$ K as a conservative benchmark in this work, while the ABRACADABRA-10 cm experiment operates at 400 mK \cite{Salemi:2021gck} and the DMRadio-$\rm{m^3}$ experiment will achieve 20 mK \cite{DMRadio:2022pkf}. Given the $\Delta t$ should be larger than the coherence time $\tau=2\pi/\Delta\omega$, we take $\Delta t = 50 \,\text{min}$ in each bandwidth scan to cover the mass range $1\times 10^{-12}\,\mathrm{eV}\sim 6\times 10^{-8}\,\mathrm{eV}$ within the total integration time of 1 year. 

Without tuning the magnetometer, we can also perform a broadband search as shown in the left plane of the Fig. \ref{setup}. In such a detection, an input coil $L_i$ inductively coupled to the SQUID magnetometer with mutual inductance $M$ connects the pickup loop and the SQUID magnetometer. Different from the resonant setup, this untunable magnetometer can access to a board band of the signal frequency over the integration time period. The magnetic flux passing through the magnetometer is $\Phi_\text{Bo}=ML^{-1}\Phi_\text{pl}$ \cite{Kahn:2016aff,Zhang:2021bpa}. 

The noise for this circuit is mainly dependent on the thermal noise felt by the magnetometer, with negligible contribution from the input coil. Corresponding to the 4 K noise temperature, at frequencies greater than $\sim4$ Hz, we expect a white flux noise floor \cite{Zhang:2021bpa}
\begin{equation}
    S_\Phi= (0.9\,\mu\Phi_0)^2/\text{Hz},
\end{equation}
with $\Phi_0 = h/(2e)=2\times 10^{-15}$ Wb is the flux quantum. When the integration time $t$ is larger than the axion coherence time $\tau$, the signal to noise ratio in the SQUID magnetometer is given by \cite{Budker:2013hfa,Kahn:2016aff}
\begin{equation}
    \frac{S}{N}=\frac{\Phi_\text{Bo}^2}{S_\Phi} \sqrt{\frac{2\pi t}{\Delta \omega}}.
\end{equation}
As a comparison, we take $t = 1$ year to cover the mass range of interest.

{\it Results} --- Finally, We perform the sensitivity estimation by setting a signal to noise ratio of 1, which gives the bounds
\begin{align}
    e_m >& 4.95\times 10^{-23} \left(\frac{m_\phi}{10^{-10}\,\mathrm{eV}}\right)^{\frac{3}{4}}   \left(\frac{0.1\,\mathrm{m}}{R}\right)^2  \left(\frac{50\,\mathrm{min}}{\Delta t}\right)^{\frac{1}{4}} \cdot \nonumber\\
    &\cdot\left(\frac{0.3 \,\mathrm{GeV/cm^3}}{\rho} \frac{5\,\mathrm{T}}{B_0} \frac{10^4}{Q}\right)^{\frac{1}{2}} \left(\frac{T}{4\,\rm K}\right)^\frac{1}{4},
\end{align}
for the toroid magnet with resonant circuit and 
\begin{align}
    e_m >& 1.73\times 10^{-23} \left(\frac{m_\phi}{10^{-10}\,\mathrm{eV}}\right)^{\frac{9}{8}}   \left(\frac{0.1\,\mathrm{m}}{R}\right)^2  \left(\frac{1\,\mathrm{yr}}{ t}\right)^{\frac{1}{8}} \cdot \nonumber\\
    &\cdot\left(\frac{0.3 \,\mathrm{GeV/cm^3}}{\rho} \frac{5\,\mathrm{T}}{B_0} \right)^{\frac{1}{2}}  \left(\frac{S_\Phi}{(0.9\,\mu\Phi_0)^2/\mathrm{Hz}}\right)^{\frac{1}{4}},
\end{align}
for the toroid magnet with broadband circuit. While that for the solenoid magnetic field can be found by the replacement $\Phi_\mathrm{pl}\to \varPhi_\mathrm{pl}$.
Here we have taken $L=1 \, \mathrm{\mu H}$ and $M= 1\, \mathrm{nH}$ as the benchmarks in both magnets to have a comparison. The constraints are obtained and presented in the $e_m/e$ versus the mass $m_\phi$ plane as shown in Fig. \ref{fig:limit}. The blue (purple) regions are excluded by a future resonant circuit detection using toroid (solenoid) magnet. The green (red) regions are disfavored by a future broadband circuit detection involving toroid (solenoid) magnet. Due to the much larger effective volume of the vector potential of the solenoid relative to that of the toroid, the bounds from the solenoid magnet can be of one order of magnitude larger than that from the solenoid magnet. It is clearly that the broadband method are more sensitive to the lower mass for the same experimental parameters with respect to the resonant method. The limits from best previous experiment and some astrophysical and cosmological observations are also shown in the same plane. In total, we find that in the mass range $1\times10^{-12}~\mathrm{eV}\sim 6 \times 10^{-8}~\mathrm{eV}$ of our interest, our results beyond the strongest existing bounds from the Red Giant \cite{Fung:2023euv} more than ten orders of magnitude at most. For the second phase with larger magnets as we mentioned before, the limits can be one to two orders of magnitude better.

We can consider simultaneous resonant and broadband detection in the future as is explored in a recently   paper \cite{Chen:2023ryb}. We can use the feedback damping circuit to achieve a high $Q$ factor while still receiving all of the signal \cite{403276,Myers2007,Kahn:2016aff,Nagahama:2016lgw,Devlin:2021fpq}. Moreover, due to the particular magnitude and direction dependence of the effective current on the vector potential background, there is room for an optimal magnet design to further enhance the power of our proposal.

\section{Acknowledgement}
This work is supported by the National Natural Science Foundation of China (NNSFC)  No. 12275134, No. 12275232, No. 12335005, and No. 12147228.

\bibliography{refs}

\onecolumngrid
\clearpage





\end{document}